\documentclass[preprint]{epl}

\begin{document}

\title{Imaging atom-clusters by hard x-ray free electron lasers
}
\author{Z. Jurek \and G. Oszl\'anyi \and G. Faigel \thanks{E-mail: \email{gf@szfki.hu}}}
\institute{Research Institute for Solid State Physics and Optics \\
           H-1525 Budapest POB. 49. Hungary}
\maketitle

\begin{abstract}
The ingenious idea of single molecule imaging by hard x-ray Free Electron Laser (X-FEL) 
pulses was recently proposed by Neutze \emph{et al.} \cite{Neutze}. However, in their numerical 
modelling of the Coulomb explosion several interactions were neglected and no reconstruction 
of the atomic structure was given. In this work we carried out improved molecular dynamics 
calculations including all quantum processes which affect the explosion. Based on this 
time evolution we generated composite elastic scattering patterns, and by using Fienup's 
algorithm successfully reconstructed the original atomic structure. The critical evaluation 
of these results gives guidelines and sets important conditions for 
future experiments aiming single molecule structure solution. 
\end{abstract}

\section{Introduction}
Structure solution without crystals was not conceivable for nearly a century. However, 
Linac based X-FEL-s will soon become operational and their extremely short and intense 
pulses offer the chance for a new type of experiment. The concept of single molecule 
imaging \cite{Neutze} is based on fast measurement of a ``slowly'' exploding system. 
In the proposed experiment the molecule does not have time to deteriorate during a single 
x-ray pulse and enough elastically scattered photons can be collected to give information on 
the unmodified structure. There are two distinct theoretical parts of the problem: the 
Coulomb explosion and the image reconstruction process. In a previous paper we have 
given a detailed description of the Coulomb explosion \cite{Jurek}. In the present work 
we analyse the effect of the explosion on the image reconstruction. 

Up to now only two papers have been published on the Coulomb explosion initiated by 
hard x-ray pulses. In the first, the explosion of a lysozime molecule has been modelled 
\cite{Neutze}. In the second, 100-1500 atom carbon clusters exploded under the influence of a
single pulse \cite{Jurek}. These works are special molecular dynamics calculations of charged particles, 
which take into account various quantum processes initiated by the photoeffect. From this 
point of view, the second set of calculations is closer to reality, because it handles all 
processes which change the dynamics of the system. Therefore, our treatment of 
Coulomb explosion is based on this work. 

Image reconstruction from intensity data - also known as the phase retrieval in optics - 
has a more extended literature. Over years of practice Fienup's hybrid input-output method 
\cite{Fienup1,Fienup2} proved to be the most successful algorithm of phase retrieval. 
It is based on a no density region surrounding a non-periodic object in real space and a corresponding 
densely sampled data in reciprocal space. Those with a crystallographic background also use the 
term \emph{oversampling} for this scenario, meaning that it uses more data than conventional 
Bragg sampling of crystals would allow \cite{Miao1}. While oversampling is not applicable in the 
case of crystals, it is well suited for single molecule imaging. A recent paper already applied 
this method for the reconstruction of 3D synthetic data of a static macromolecule \cite{Miao2}. 

None of the earlier papers treated the Coulomb explosion and the reconstruction process 
together, which we shall do in the present work. We also discuss the fundamental 
questions: how the atom density affects the Coulomb explosion and reconstruction, what 
time-window of a single pulse is available for useful data collection and finally how 
multi-pulse measurements affect reconstruction. The answers are crucial for planning an 
experiment with the target of single molecule structure solution. 

\section{Coulomb explosion of the cluster}
Atomic resolution imaging is important in many fields of science, from solid state 
physics to biology. This diverse interest covers cluster sizes ranging from a few tens to 
millions of atoms. The original proposal of single molecule imaging came from biology 
but biological macromolecules are too large to accurately model their Coulomb explosion. 
Therefore, we follow the time evolution of smaller atomic clusters. We expect that the 
results can be plausibly extended to larger systems and the general conclusions are also 
relevant to macromolecules.

Our model system is a 200-atom carbon cluster forming an incomplete cube, as shown in 
fig.~\ref{fig1}. Atoms are scattered around grid points of a simple cubic lattice and are held 
together by central forces. Three different lattice spacings were investigated: $a$=1.5, 2.4 and 
3.0~{\AA}. The depth of the bonding potential corresponds to the covalent bond for 
the 1.5~{\AA} lattice, to the van der Waals bond for the 3.0~{\AA} lattice and it is linearly 
interpolated for the intermediate 2.4~{\AA} lattice. With this choice of parameters we can 
study the characteristics of the explosion for a wide range of atom density and bond 
strength.

The calculation follows the time evolution of the system while a gaussian shape x-ray 
pulse is incident on the sample. We solve the classical equations of motion for all the 
particles (atoms, electrons and ions) as described in \cite{Jurek}. Quantum processes are taken 
into account as random events by probabilities corresponding to their cross sections. We 
include the photoeffect, Auger processes, Coulomb interaction and inelastic electron scattering. 
The parameters of the pulse are: 10 keV photon energy, 
$5\cdot 10^{12}$ integrated photon number and 10~fs full width at half maximum. Note that the 
pulse width is shorter than initially expected at hard x-ray FEL-s \cite{Materlik,Arthur}. 
However, current efforts to decrease the pulse width below 100~fs are promising and reaching 10~fs 
is not unrealistic \cite{Wang}. The reason for this choice is that the shorter the pulse 
the higher the ratio of photons available for useful imaging. 

To illustrate the changes of the atomic configuration during a pulse, the time evolution of 
a typical Coulomb explosion of the $a$=1.5~{\AA} cluster is shown in fig.~\ref{fig2}. The total 
period of model calculation is between $t$=0 and 30~fs where 15~fs corresponds to the maximum 
of the gaussian shape pulse. It is clear that up to 10~fs the cluster is almost unchanged. 
Visible distortions start to develop at 12~fs and by 18~fs the cluster has totally 
disintegrated. To quantify the explosion we calculated the cluster charge and the 
mean displacement as a function of time for the three atom densities (see fig.~\ref{fig3}). 
The total charge shows how many electrons scatter elastically while the mean displacement 
gives an estimate for the distortion of geometry. The trend in both series of curves 
indicates that the explosion takes longer for low atom densities. This is in agreement with 
natural expectations, the Coulomb repulsion is smaller for ions at larger distances. We 
can learn two important facts from figs.~\ref{fig2} and \ref{fig3}. 
(i) useful data can be collected only in the first part of the pulse. 
(ii) low density systems will allow more time for data collection. 

However, no cleverly chosen parameter can replace the real reconstruction process. 
During the explosion the elastic scattering pattern changes significantly and what we 
measure is a composite pattern. It is not trivial that using this composite pattern as input 
data, the original atomic configuration can be reconstructed. 

\section{The reconstruction algorithm}
In this work we used a modification of Fienup's hybrid input-output algorithm. In real 
space the charge density is represented in a cubic box with cell edge $L$ and grid spacing 
$\Delta L$. The box size $L$ must be larger than the object and the grid spacing $\Delta L$ must be 
sufficiently fine to reconstruct atoms. We used $L$=25, 35 and 40~{\AA} for the three different 
atom densities while $\Delta L$=0.4~{\AA} was kept constant. The real space charge density and the 
reciprocal space scattering amplitudes are related by the Discrete Fourier Transform. 
Thus amplitudes in reciprocal space are also represented in a cubic box with a cell edge 
$1/\Delta L$ and grid spacing $1/L$. Measurement of elastic scattering is required at points 
of this dense grid. The reciprocal space box limits the maximum momentum transfer at 
$q_\mathrm{max}=\pi/\Delta L$. We use all data within the sphere of radius $q_\mathrm{max}$ 
and treat unobserved data  outside the sphere as zeros. 

The algorithm requires a molecular support, which completely surrounds the object. The 
object is smaller than its support and the charge density is positive - these are the simple a 
priori constraints. We used a spherical volume, which is a loose support, it poorly approximates 
the shape of the cluster. The volume ratio $V_\mathrm{box}/V_\mathrm{support}$ is called the 
oversampling parameter, denoted by $\sigma$. For the three atom densities we chose the radius 
10, 14 and 16~{\AA}, corresponding to $\sigma=3.73$ for which the algorithm is expected to work. 
Fienup's method is a special type of density modification, it cycles between real 
and reciprocal space by the Fourier transform. In real space the charge density is 
modified only where the support or positivity constraints are violated, while in reciprocal 
space the observed moduli and model phases are combined without a condition. The $q=0$ 
reciprocal space amplitude corresponds to the total charge but is unobserved. We chose 
the correct ab initio treatment, it is initialized to zero and then allowed to change freely. 
The iteration starts with a random phase set in reciprocal space. Testing multiple phase 
sets is important, some phase sets converge faster, while others do not converge within a 
fixed number of iterations. There is a single parameter $\beta$ in the real part of the algorithm, 
which acts as feedback. In the literature the typical value of $\beta$ is 0.5-0.9, we used $\beta=0.7$. 
Our experience with the original algorithm is that after a number of iterations large 
negative charge density develops and oscillates which is a sign of over-amplification. The 
solution was to set the charge density outside the support to zero before each real space 
modification. This makes convergence slower but the quality of reconstruction is better. 
To evaluate a solution it is best both to plot the charge density and by peak picking 
analyse the number and position of atoms.

Our model cluster is a non-centrosymmetric object with pseudo-symmetries, its atomic 
resolution reconstruction is far from trivial. We carefully tested the algorithm by 
reconstructing static configurations of the exploding cluster at different times up to 20~fs. 
We found that any static configuration can be reconstructed in this time interval, so only 
the quality of real data will limit the structure solution. 

\section{3D data and multi-pulse measurements}

The reconstruction algorithm needs 3D data while in a single-pulse experiment only a 2D 
slice can be collected. Therefore, even at ideal conditions, we cannot measure the full 
data set during a single pulse. We have to merge the results of many repeated 
measurements, which are made on identical replicas of the molecule in different 
orientations. Here we give an estimate for the minimum number of independent 
orientations and the total number of pulses necessary for a complete data set from a small 
biological molecule. 

Let us take a typical macromolecule which is enclosed in the real space box with edge 
$L$=60~{\AA} and require reconstruction on a grid with $\Delta L$=0.4~{\AA} spacing. Then reciprocal 
space data is represented in a box with edge $1/\Delta L$ and grid spacing $1/L$. At the maximum 
resolution data is accessible only inside the limiting sphere with radius $R=1/(2\Delta L)$. The 
scattering intensity is sampled with $(1/L)^3$ density and $4\pi /3\cdot R^3\cdot L^3=1.8\cdot 10^6$
is the total number of 3D data points. A single orientation provides 2D data on the surface of the 
Ewald sphere with radius $R/2$. Assuming that each $(1/L)^2$ area of the $4\pi (R/2)^2$ surface 
gives independent information, a single orientation contributes $1.8\cdot 10^4$ points to the 3D 
data. So in this example the number of independent orientations is 100. For larger 
molecules this number increases with the linear size of the box.

This is the minimum number of orientations, which must be measured with adequate 
statistics. Repeated measurements are needed both to cover all required orientations and 
also to improve their statistics. Sorting individual single-pulse patterns into orientation 
bins is not discussed here but will need special attention. Let us suppose there is no 
background, and the reconstruction algorithm can tolerate 10\% noise. Considering that x-
ray scattering is strongly anisotropic we estimate that on average 100 elastically scattered 
photons per pixel must be collected. For the above example this is the order of $10^8$ 
photons/orientation and $10^{10}$ photons total. The elastic cross section of carbon and the 
intensity of $5\cdot 10^{12}$ photons/pulse lead to about $5\cdot 10^3$ elastically scattered photons
per pulse. Therefore, to construct 3D data with adequate statistics approximately $2\cdot 10^6$ 
repeated measurements must be done using the same number of replicas of the molecule. 

\section{Time integral and gating}
The Coulomb explosion fundamentally determines the quality of data, which is available 
for structure solution. During data collection the scattering pattern changes drastically. 
Even a 2D section of data measured by a single pulse is a composite pattern, a time 
integral weighted by the pulse intensity. 

We need the optimum integration time with the highest number of elastically scattered 
photons and a pattern closest to that of the original structure. If the time-window is too 
short then the statistics of individual patterns is poor. If the time-window is too long then 
the scattering of the disintegrated cluster becomes dominant, washing out the original 
structural information. For the sake of data quality part of the pulse must be thrown away 
and because the total number of photons is fixed, more measurements are needed for the 
same statistics. We will show that with the pulse width of 10~fs the increase in the 
number of measurements is not drastic, a factor of 2 to 4. The situation could be worse 
with the pulse width of 100~fs because the same level of distortion is reached much 
before the peak, which means an order of magnitude less photons for reconstruction. 

Here we give an estimate for the optimum integration time using 10~fs pulses. For 
simplicity we assume that 3D data can be obtained from a single experiment, as if 
separate explosions took place the same way. Integration is approximated by summing 
the scattering pattern at each 0.5 fs time-slice with the weight of the gaussian pulse intensity. 
The upper limit of integration is $t_\mathrm{max}$. Statistical uncertainty of real experiments 
is simulated by adding 10\% noise to the composite pattern. Then we make a reconstruction for each 
composite pattern and compare it to the original charge density. 
We accept the reconstruction for $t_\mathrm{max}$ if all the atoms are found and the standard 
deviation of displacement does not exceed 0.3~{\AA}. With these requirements the structure of the 
$a$=1.5, 2.4 and 3.0~{\AA} cluster was successfully recovered up to $t_\mathrm{max}$=12, 14 and 15~fs 
respectively. These time-windows correspond to 25, 40 and 50\% useful photons per pulse \cite{tmax}.

Collecting data for longer periods degrades the quality of the reconstruction, the useful 
structural information is quickly washed out. It is informative to follow the scattering 
pattern of individual time-slices. What happens is that the disintegrating cluster becomes 
random in real space, so its Fourier transform turns into white noise. Fig.~\ref{fig4} shows the 
quality of reconstruction for the $a$=1.5~{\AA} cluster using integrated data up to $t_\mathrm{max}$=12 
and 18~fs. While with shorter integration we get a structure practically identical to the original 
cluster, just 6~fs longer integration leads to the loss of half of the atoms and to increased 
structural disorder. Similar differences were found in the case of clusters with larger 
interatomic distances. The summary of the above results is: accurate gating will be 
critical for a real experiment. 

\section{The effect of independent replicas}
After giving an estimate for the optimum integration time of single pulses, we consider 
the effect of averaging in multi-pulse experiments. Although in principle identical 
replicas of the molecule are used in consecutive experiments, their disintegration will 
follow different pathways. This is partly due to thermal vibration which displaces the 
starting atomic positions, and partly to the stochastic nature of photo-ionization, Auger 
process and secondary ionization which changes the dynamics of the Coulomb explosion. 

To check the joint effect of the time integral and multi-pulse averaging on structure 
solution, we followed 100 explosions of the $a$=1.5~{\AA} cluster. The deviation of the starting 
configurations was chosen to be small, corresponding to low temperature (T=30 K). Even 
so, the time evolution leads to differences in the atomic position and charge state which is 
already apparent at $t_\mathrm{max}$=12~fs. The scattering pattern was calculated as before, but 
the time integral was added for all individual explosions. We found that the quality of 
reconstruction is somewhat better than for the single-pulse case. This is in accordance 
with expectations, averaging the contribution of many slightly different 
configurations results in a smaller deviation from the original structure. The improvement 
is not drastic, the useful integration time can be increased just by a femtosecond. 

\section{Conclusion and outlook}
In this work we have shown that single molecule imaging using hard X-FEL pulses might 
be possible, but the experimental difficulties are more severe than anticipated earlier. Our 
results differ from those of previous publications in several respects. First, the time 
evolution of the Coulomb explosion is more reliable as we included all quantum 
processes in the simulation. Second, we time integrated the scattering pattern of the 
exploding cluster and took into account the effect of multiple replicas. Reconstruction of 
the original atomic structure was achieved using these composite patterns at 0.8~{\AA} 
resolution. This is in contrast to previous work, which reconstructed a static structure at 
2.5~{\AA} resolution \cite{Miao2}.

While giving a unified treatment of the Coulomb explosion and reconstruction, we 
arrived at new conclusions: (i) We demonstrated that only part of the pulses can be 
efficiently used for data collection, therefore very fast gating will be required. (ii) We 
analysed the reconstruction as a function of atom density, and found that lower density 
will allow more time for data collection. In terms of the fraction of total photons available 
for imaging this means a maximum of 50\%. (iii) We also showed that averaging the 
contribution of individual replicas in a multi-pulse experiment slightly improves the 
quality of reconstruction.

Finally, we would like to call attention to problems not discussed in the paper, but 
important for the feasibility of real experiments. In practice a complete data set can be 
collected by measuring a large number of identical replicas, and individual scattering 
patterns have to be arranged into 3D data on a reciprocal space grid. Due to the low 
statistics of single patterns it will be difficult to sort them into distinct orientation bins. 
Although electron microscopy successfully handles a similar problem \cite{EM}, in the case of 
X-FEL experiments it remains to be solved. An other serious problem is the low signal to 
noise ratio. As energy selective detectors are useless at the femtosecond timescale, the 
detector will count all the photons in the experiment. As the elastic scattering is weak, 
any spurious scattering will distort useful structural information. It is difficult to estimate 
the background level, but Compton scattering will certainly give a contribution. Its 
weight relative to elastic scattering is small while the electrons are localized on the 
atoms, but will drastically increase after photoionization. Of course, background will also 
come from instrumental sources such as the beam path and sample environment. Here we 
pointed out just a few experimental problems, probably there will be several others. 
However, the importance of single molecule imaging is so great that once it is shown to 
work given ideal conditions, it is worth the effort to realize the experiment and attempt 
the solution of these problems.

\acknowledgments
We thank J\'anos Hajd\'u and Abraham Sz\H oke for the illuminating discussions. This 
research was funded by OTKA grants T043237 and T043494 and G.O. was also 
supported by a Bolyai J\'anos Scholarship.

\begin{figure}
\caption{Model system of the 200 atom carbon cluster forming an incomplete cube.
Only the most dense variant with 1.5~{\AA} interatomic distances is shown.}
\label{fig1}
\end{figure}

\begin{figure}
\caption{Time evolution of a typical Coulomb explosion of the $a$=1.5~{\AA} cluster.
The four snapshots are taken at $t$=10, 12, 15 and 18~fs, where t=15~fs corresponds to the 
maximum of the x-ray pulse.}
\label{fig2}
\end{figure}

\begin{figure}
\onefigure[width=6cm]{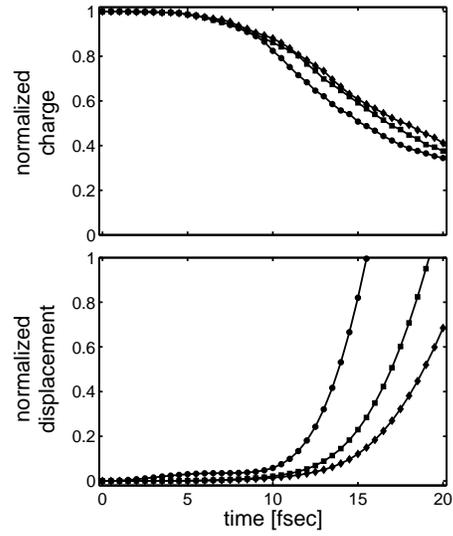}
\caption{Time evolution of the cluster charge and mean displacement for the three atom densities.
The charge is normalized to $1$, the mean displacement is scaled by $1/a$. The symbols 
circle, square and diamond correspond to nearest atom distances $a$=1.5, 2.4 and 3.0~{\AA} respectively.}
\label{fig3}
\end{figure}

\begin{figure}
\caption{Reconstructed charge density of the $a$=1.5~{\AA} cluster using integrated data
up to $t_{max}$=12 and 18~fs. The difference of quality is obvious.}
\label{fig4}
\end{figure}


\begin{thebibliography}{0}
\bibitem{tmax}
  \Page{The integration limits here differ from those of our earlier publication. 
        The reason is that our previous estimate was simply based on geometry and charge state,
        without generating composite scattering patterns and attempting reconstruction.}
\end{thebibliography}
\end{document}